\documentclass[debug,overfull]{epl} 
\usepackage{graphicx,wrapfig}
\usepackage{dcolumn}
\usepackage{amsmath}
\usepackage[latin1]{inputenc} 
\title{Log-Poisson statistics and full aging in glassy systems}
\shorttitle{Log-Poisson statistics and full aging}
\author{Paolo Sibani \and Jesper Dall} 
\institute{Fysisk Institut, 
Syddansk Universitet--5230 Odense M, Denmark}
\begin{document}
\maketitle

\begin{abstract}
\noindent We argue that Poisson statistics in logarithmic
time provides an idealized description of
non-equilibrium configurational rearrangements in 
aging glassy systems. The description puts stringent
requirements on the geometry of the metastable attractors
visited at age $t_w$.
Analytical implications for the residence time distributions
as function of $t_w$ and the correlation functions are
derived. These are verified by  extensive numerical studies  
of short range Ising spin glasses.  
\end{abstract}

\section{Introduction} \label{introduction}
Aging is a key aspect of the dynamics in a host of complex glassy 
systems~\cite{Reim86,Refrigier87c,Andersson92,Nicodemi01,Bureau02,Hannemann02,Kityk02,Bissig03}. 
Its main characteristics are a slow, sub-linear drift of macroscopic averages 
with the age $t_w$ of the system, i.e.\ the time elapsed after an initial 
thermal quench, and a concomitant age dependence of the correlation and response
functions. The latter shows that a quasi-stationary fluctuation regime 
for short observation times $t < t_w$ is followed by 
off-equilibrium dynamics for $t > t_w$. 
This behavior can be attributed to an increasing degree of thermal stability 
of the dynamically relevant attractors or landscape valleys
as the age $t_w$ grows~\cite{Sibani89,Hoffmann90,Vincent95,Bouchaud92}.

Within each valley, the pseudo-equilibrium fluctuations 
are determined by the free energy.  
However, the inter-valley non-equilibrium dynamics
could be practically  irreversible, and hence 
qualitatively  different.
Adopting   the simplified view that the selection is 
truly irreversible  allows us to formulate a 
theoretical description of large non-equilibrium rearrangements,
or `quakes', which borrows 
from studies of non-thermal models~\cite{Sibani93a,Sibani01}
and which supplies  a  basis
to the striking similarity of complex  dynamics
seen in very different  systems with multiple metastable states.
The key feature is dynamical invariance to shifts 
in the \emph{logarithm} of time. This implies
that the time spent in thermally metastable 
attractors is  power-law distributed with an exponent $\alpha$.
Furthermore, the distribution has  a  $t/t_w$ aging behavior, often called 
full or pure aging. The exponent $\alpha$ depends on the system
size but not on the temperature. The further assumption 
that the memory of the initial configuration decays as $\exp(-\gamma k)$
with the number $k$ of quakes   leads to  
a  power-law decay and a $t/t_w$ form  of the correlation 
function for $t>t_w$. The corresponding  
non-equilibrium exponent $\lambda$ is 
jointly determined by $\alpha$ and $\gamma$. 

These  theoretical 
results  are checked   through  extensive numerical 
simulations of short ranged spin glass systems. 
While direct experimental evidence is not yet available, possibly related 
large-scale irreversible rearrangements similar to quakes  have recently 
been identified in aging colloidal gels~\cite{Bissig03}.

\section{Basic formalism} 
Beyond its  original macro-evolutionary 
context, the concept of `punctuated equilibrium'~\cite{Gould77} 
is   applicable to any stochastic dynamics where long periods of stagnation
are dotted by fast irreversible changes. This occurs e.g.\ in 
micro-evolution~\cite{Lenski94,Sibani99a} 
and in certain driven dissipative models~\cite{Sibani93a,Sibani01}.
Within a population evolving in a fitness landscape~\cite{Sibani99a},
fitness \emph{records} achieved by an individual through random 
mutations are amplified to the macroscopic level by 
darwinistic selection and trigger punctuations. Since the same statistics 
describes both punctuations and records, the
analytical description available for the latter~\cite{Sibani93a} can 
be applied to the former.
Below, we discuss how thermal noise 
may similarly induce large and irreversible quakes in glassy systems.
 
To summarize the basic ideas of record  statistics~\cite{Sibani93a}, 
consider times $t_w \geq 1$ 
and $t \geq 0$, and independently draw for each integer time a 
random number from any fixed distribution not supported on a finite set.
The attempts whose outcome exceeds  all previous outcomes
are marked as records. For large times, the probability 
that $k$ such events occur in $[t_w,t_w+t]$ is then 
\begin{equation}
P_k(t_w,t_w + t) = 
\frac{1}{k!} 
\left[ \alpha \log \left( \frac{t_w + t}{t_w} \right) \right]^k 
\left[ \frac{t_w + t}{t_w} \right]^{-\alpha},
\label{logPoisson}
\end{equation} 
where    $\alpha =1$ for a single sequence of 
attempts. The total number of records 
produced by $n$ independent sequences  is a sum of log-Poisson
variables, and hence itself log-Poisson distributed, with $\alpha=n$. 

Equation~(\ref{logPoisson}) can be recognized as a Poisson distribution  
with $\log(t_w + t )- \log(t_w)$ replacing the 
linear difference $t$ between the time arguments, or, for short, 
a log-Poisson distribution\footnote{ 
The equation  was derived in~\cite{Sibani93a} for $t_w=1$, but the 
generalization to the propagator $P_n(t_w,t_w + t)$ 
is straightforward and completely analogous to the standard case.}.
Equation~(\ref{logPoisson}) implies that the rate of events decays 
as $\alpha/(t+t_w)$, while the logarithmic rate of events approaches 
the constant value $\alpha$ for $t \gg t_w$. 
\begin{figure}[ht!]
\twofigures[scale=0.37]{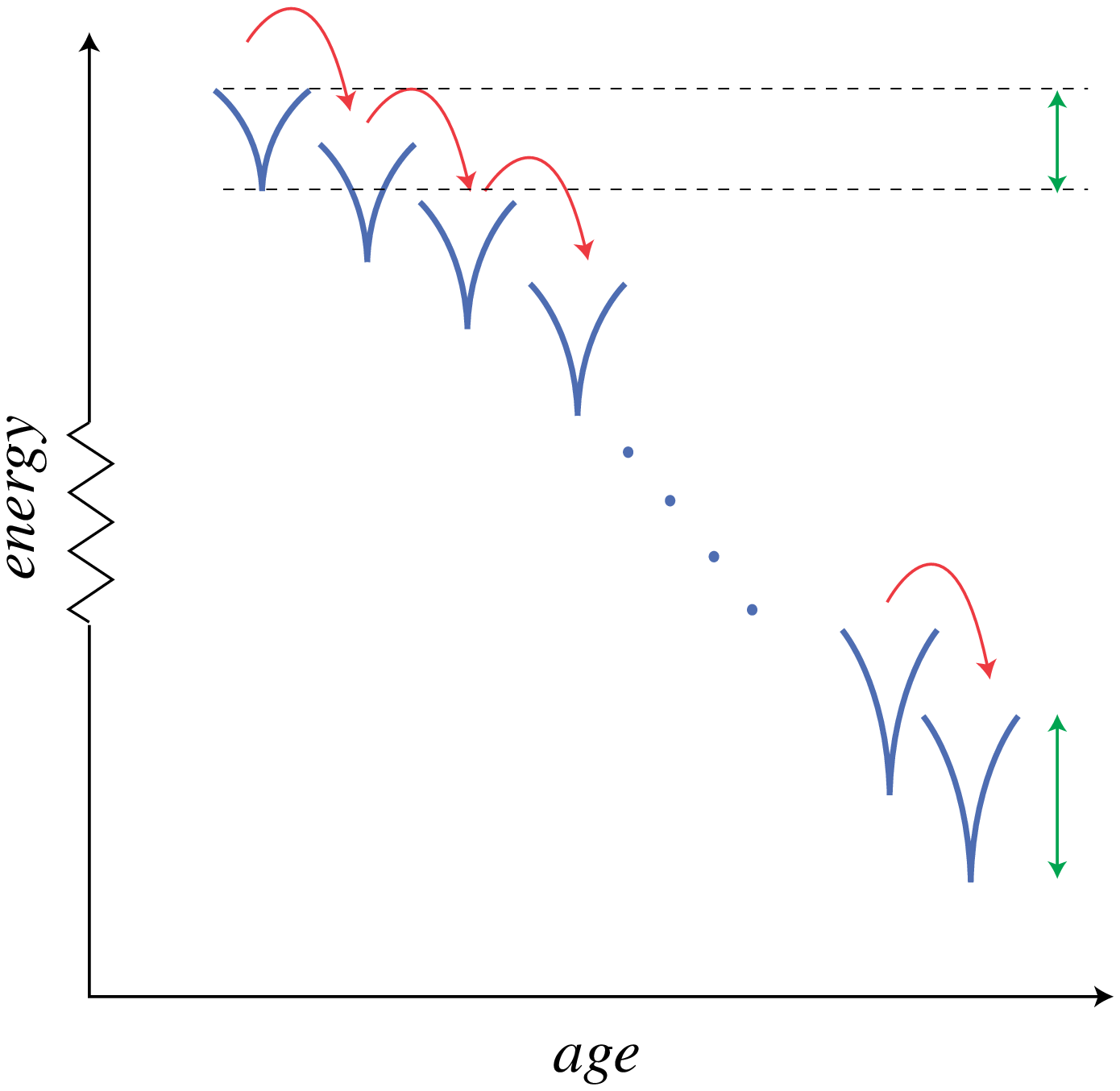}{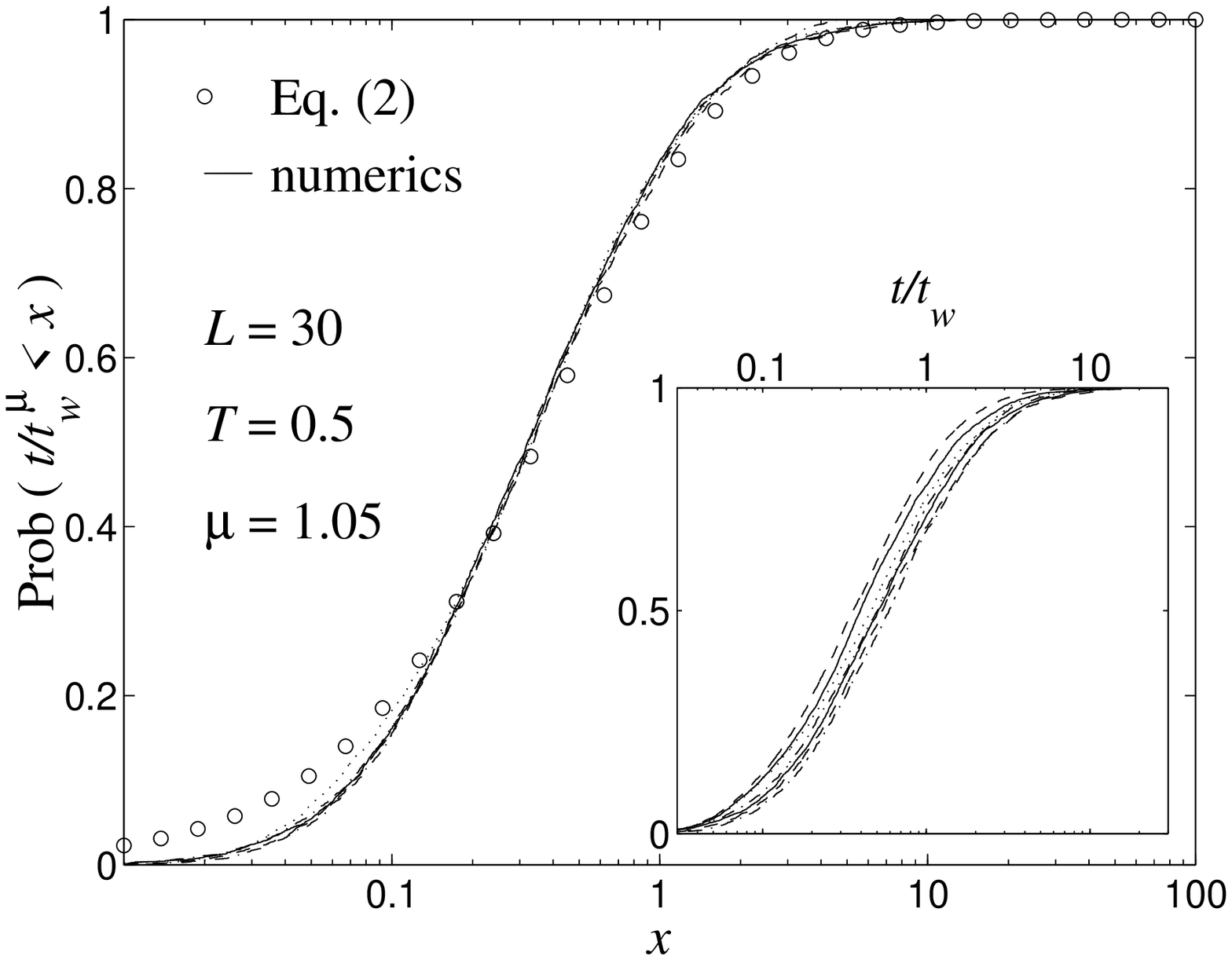}
\caption
{ \small 
An idealized  landscape where the attractors are symbolized by wedges.
The arrows represent `quakes' leading from one attractor to the next. 
They entail a large energy change---the vertical displacement---and 
are therefore irreversible. The  log-Poisson statistics of the quakes
originates from the infinitesimal increase of the  metastability 
of the attractors selected, which is rendered by a slight increase in the 
bottom to top distance (depth) of each trap. 
}
\label{land_sketch}
\caption{ \small 
The shown curves  all describe  the distribution of 
a suitably rescaled residence time in the 
attractors---or valleys---of the sort 
sketched in   Fig.~\ref{land_sketch}.
The  circles  are the prediction of the log-Poisson
theory---Eqs.~(\ref{logPoisson}) and~(\ref{an_scaling})---and  
correspond to  full aging $(\mu=1)$. 
 The empirical data  for the spin glass landscape
are plotted as lines, and describe the 
  scaled residence time $t_r/t_w^\mu$ in valleys entered  
at $10^2 \leq t_w \leq 10^5$. They merge into 
a seemingly continuous line due to the quality of the data collapse.
The insert shows the same data with the less satisfactory
$\mu=1$ scaling. 
}
\label{residence_t}
\end{figure}

Glossing over the discreteness of the time arguments,
we can now pick  thermal  noise  
as the  (arbitrary) source of randomness   producing the records. 
Since landscape induced correlations are negligible if records are intercalated  
by long intervals spent dwelling near a local energy minimum, 
the  number of energy records
observed in a fixed time interval is expected to be
 proportional to a log-Poisson variable~\cite{Dall03}.
The same applies if energies are measured relative to the current
 lowest energy minimum, producing the  energy difference
 henceforth   referred to  as a `barrier'. With this
 definition, not all  barrier crossings   carry  physical 
significance. However,  if quakes  are   log-Poisson distributed, 
they are  naturally  associated  with 
the  crossing  of   barriers of record magnitude.  
Two  special landscape  features  sketched   in Fig.\ref{land_sketch}
would allow barrier records to induce quakes: 
i) The stability of the
attractors successively selected, as gauged by  e.g.  an exit barrier,
must only increase  marginally~\cite{Sibani93a,Sibani01}, 
i.e.\ in principle by an  infinitesimal amount.
 This requires that   shallow attractors be predominant, and   
 implies that   the  actual  magnitude  of the records generating
 the quakes is immaterial\footnote{The predominance of shallow attractors concurs with
  the observation that thermal quenches
produce shallow minima in glassy systems, and fits  the 
anomalous  sensitivity of complex relaxation to small mechanical or
thermal perturbations implied by the so-called reset or memory effects~\cite{Berthier02}.}.
ii) A large   and  negative energy change associated to
a quake, ensures its irreversibility. Such  changes 
are  observed  in both atomistic glass models~\cite{Schon00} 
and spin glasses~\cite{Dall03},
 
The only adjustable quantities  in this 
 landscape  cartoon   are the parameter $\alpha$  and 
 the form of the quake induced decorrelation. 
The noise independence of the statistics restricts $\alpha$ to be a 
temperature independent quantity. 
An interesting increase with system size  can arise 
in spatially extended systems, where  sufficiently  distant regions  
can evolve  independently.   

In the sequel, we first deal with some theoretical  consequences of 
the marginal stability of the attractors
and irreversibility of the quakes. 
To make contact with  experiments,
 we then   summarize an empirical way  to identify quakes in
glassy systems~\cite{Dall03} and  apply it to spin glasses.  

\section{Attractor residence times}
The distribution of the residence or trapping time $t_r$ in valleys of   
an energy landscape has   been  considered theoretically~\cite{Vincent95,Bouchaud92}, but
has not previously been  measured. 

Let $t_k$ mark the occurrence of the $k$'th event 
in a log-Poisson process with logarithmic rate $\alpha$. 
In perfect analogy to the usual Poisson process, the `log-waiting time' 
$\log(t_k) - \log(t_{k-1})$  has the exponential distribution 
\[ 
{\rm Prob} ( \log(t_k/t_{k-1}) < x ) = 1 - \exp(-\alpha x),
\] 
which is  mathematically equivalent to the   event 
distribution given in Eq.~(\ref{logPoisson}).
We now consider the distribution of the residence time 
$t_r= t_k - t_{k-1}$ for traps entered at $ t_{k-1} = t_w$.
As $t_r/t_w = t_k/t_{k-1} - 1$, we find  
\begin{equation}
{\rm Prob}( t_r/t_w < x )= 1 - (x+1)^{-\alpha},
\label{an_scaling}
\end{equation}
which corresponds to  full aging behavior.
Equation.~(\ref{an_scaling}) with $\alpha = 2.3$ is plotted
(circles) in the main panel of Fig.~\ref{residence_t},
together with spin glass data (lines). These  are   scaled in a slightly different
manner,  to account for  a small but systematic
deviation from full  aging.   

The power-law form of Eq.~(\ref{an_scaling}) was suggested by 
Bouchaud~\cite{Bouchaud92}, with the added and crucial assumption of 
an infinite average residence time. In our case, the average residence time 
$\langle t_r \rangle$ is equal to 
$t_w/(\alpha - 1)$ which is finite and $\approx t_w$ for the observed  
$\alpha\approx 2$. As a   consistency  
check, we  note that the largest 
relaxation time $\tau_{eq}$ in a valley
must be smaller than  
$\langle t_r\rangle$, for 
 local equilibration to apply.  When the latter  is of order $t_w$,
   $\tau_{eq} < t_w$ follows.
In the lack of special symmetries~\cite{VanKampen92},
$\tau_{eq} = \tau_{corr} $, where $\tau_{corr} $ characterizes the short time decay 
of correlation functions, either as a cut-off parameter or 
directly as the time constant of an exponential decay. Hence $\tau_{corr} < t_w$   
as  expected.

\section{Scaling of correlations} 
Additional information on the effect of the quakes enters
the calculation of  thermally averaged correlations functions. 
Let $c(m;m+k)$ be the configurational overlap between 
the lowest energy configurations in valley  
$m$ and $m+k$. With probability $P_m(1,t_w)P_k(t_w,t_w+t)$,
$m$ and $m+k$ are the `current' attractors at times 
$t_w$ and $t_w+t$, whence the \emph{non}-equilibrium part 
of the configurational autocorrelation function is given by
\begin{equation}
\overline{C}(t_w, t_w+t) = 
\sum_{m=1}^\infty \sum_{k=0}^\infty c(m;m+k) P_m(1,t_w) P_k(t_w,t_w+t).
\label{two-point}
\end{equation} 

Of special interest is $c(1;1+k)$, the average overlap between the first and 
the ($1+k)$'th valley. This is the sole contribution to
Eq.~(\ref{two-point}) for $ t_w=1 $, in which case  $\overline{C}(1,1+t)$ 
additionally describes the  magnetization decay of
the fully polarized  configuration of a spin glass~\cite{Kisker96}. 
The exponential form 
\begin{equation}
c(1;1+k) = (1-c_\infty) e^{-\gamma(T) k} + c_\infty
\label{twoparameterfit}
\end{equation}
anticipates the results in Fig.~\ref{correlation}.  
An exponential decay means that each event induces the 
same relative change, and leads to 
\begin{equation}
\overline{C}(1,1+t) = 
\sum_{ k=0}^\infty c(1;1+k) P_k(1,1+t) = (1-c_\infty)  t^{-\lambda} + c_\infty, 
\label{corre}
\end{equation} 
where the non-equilibrium exponent $\lambda(T)$ is given by 
\begin{equation}
\lambda(T) = \alpha (1-e^{-\gamma(T)}) \approx \alpha \gamma(T).
\label{expos}
\end{equation}
Equation~(\ref{expos}) thus links a dynamical exponent to the landscape 
geometry. The second approximate equality given in Eq.~(\ref{expos})
is only justified for $\gamma \ll 1$.

The further assumption that $c(m;m+k)$ is 
altogether independent of $m$ amounts to translational invariance
with respect to $m$, and hence with respect to additive shifts 
in the logarithm of $t_w$. These are equivalent to a multiplicative
rescaling of $t_w$. Not surprisingly, 
 Eq.~(\ref{two-point}) yields  the scale invariant  and full aging   form 
\begin{equation}
\overline{C}(t_w, t_w+t) \propto 
\left( \frac{t+t_w}{t_w} \right) ^{-\lambda}.
\label{simple_decorr} 
\end{equation}

\section{Landscape exploration} \label{landscape}
How should valleys and quakes be concretely defined given a time 
sequence of data? One approach is the protocol developed~\cite{Dall03}
for numerical data describing isothermal relaxation after a deep quench.
The basic observation is that at sufficiently low $T$ pseudo-equilibrium fluctuations 
repeatedly visit the same local energy minimum. Hence, the 
attainment of an energy value lower than all previous values 
implies a non-equilibrium event, i.e.\ an ongoing quake.  
Several closely spaced  minima records will often  occur as a  quake
 slides downhill among which  
  the   one  preceding a barrier record
is likely to be the  most physically significant,
as it---ideally---coincides with a new local minimum. 
By the same token, in a   series
of   barrier records corresponding to  
  an uphill climb, the last record 
preceding a low energy record is  
  most significant, as it marks the access to a new local 
  minimum, and hence the start of 
  a quake. The times at which the above 
  types of event  occur provide an empirical definition of 
  both valleys and quakes. 

This procedure yields a null result if i) only a single energy minimum 
is present; or ii) the initial condition is the ground state irrespective 
of the number of local minima present; or iii) all the minima and/or barriers 
are equal. Also, iv) if the trajectories can bypass the energy barriers and 
edge their way into states of lower energy through `entropic barriers', 
the number of valleys discovered remains less or equal to one at all times.
  
The actual outcome   is very different, as illustrated in Fig.~\ref{no_valleys}, 
where the estimated  average and variance of the number $n_V$ of valleys visited 
is plotted versus the logarithm of time. An equality between 
variance and average is a key feature of Poisson distributions.
If we neglect the initial curvature, the estimated quantities are very 
nearly proportional to $\log t$, indicating that $n_V$ itself is proportional 
to a log-Poisson variable through a constant of order one.  
We emphasize that log-Poisson statistics is an idealized low $T$ description,   
and that systematic corrections in 
the form of a decaying logarithmic rate of events must be expected, since 
each valley has a growing probability to contain the true ground 
state. Since $\alpha$ is not truly constant, 
the single value required by the log-Poisson theory cannot be estimated accurately.  
 
As $T$ grows, local thermal equilibration within each attractor  
is not granted, and low energy records lose  connection to landscape minima. 
A check of the energy and configurational changes associated 
with the quakes, of the sort performed in~\cite{Dall03} is required 
to independently establish their physical relevance.   
We additionally checked that the  low energy record configurations are 
close (albeit usually not identical) to actual valley minima.

\section{Numerical results} \label{results}
We performed extensive simulations of short-ranged Ising spin glasses on cubic 
lattices ($L^3$ spins) with periodic boundary conditions and symmetric Gaussian 
couplings with unit variance, see also~\cite{Dall03}. 
The Waiting Time Method~\cite{Dall01} with 
single spin flip dynamics, which is dynamically equivalent to the Metropolis 
algorithm but much faster at low $T$, was used.  
The `intrinsic' and size independent time variable of this method
corresponds to the number of Monte Carlo (lattice) sweeps in the Metropolis 
as well as to the physical time of a real experiment.
In all our runs, we skip the data within the first $10$
time units in order to let the system settle down from 
its random $T = \infty$ initial configuration. 
 
\begin{figure} 
\twofigures[scale=0.37]{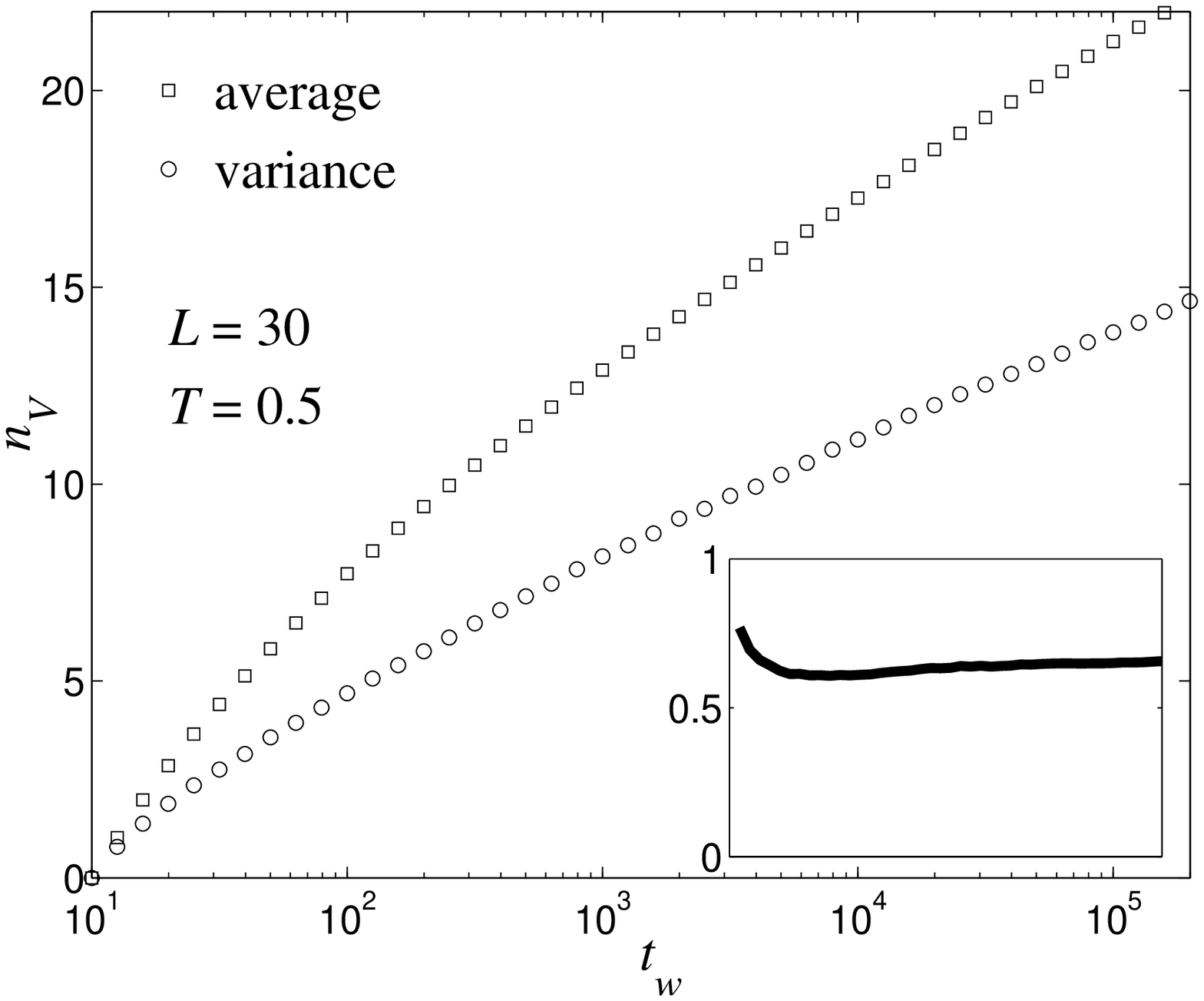}{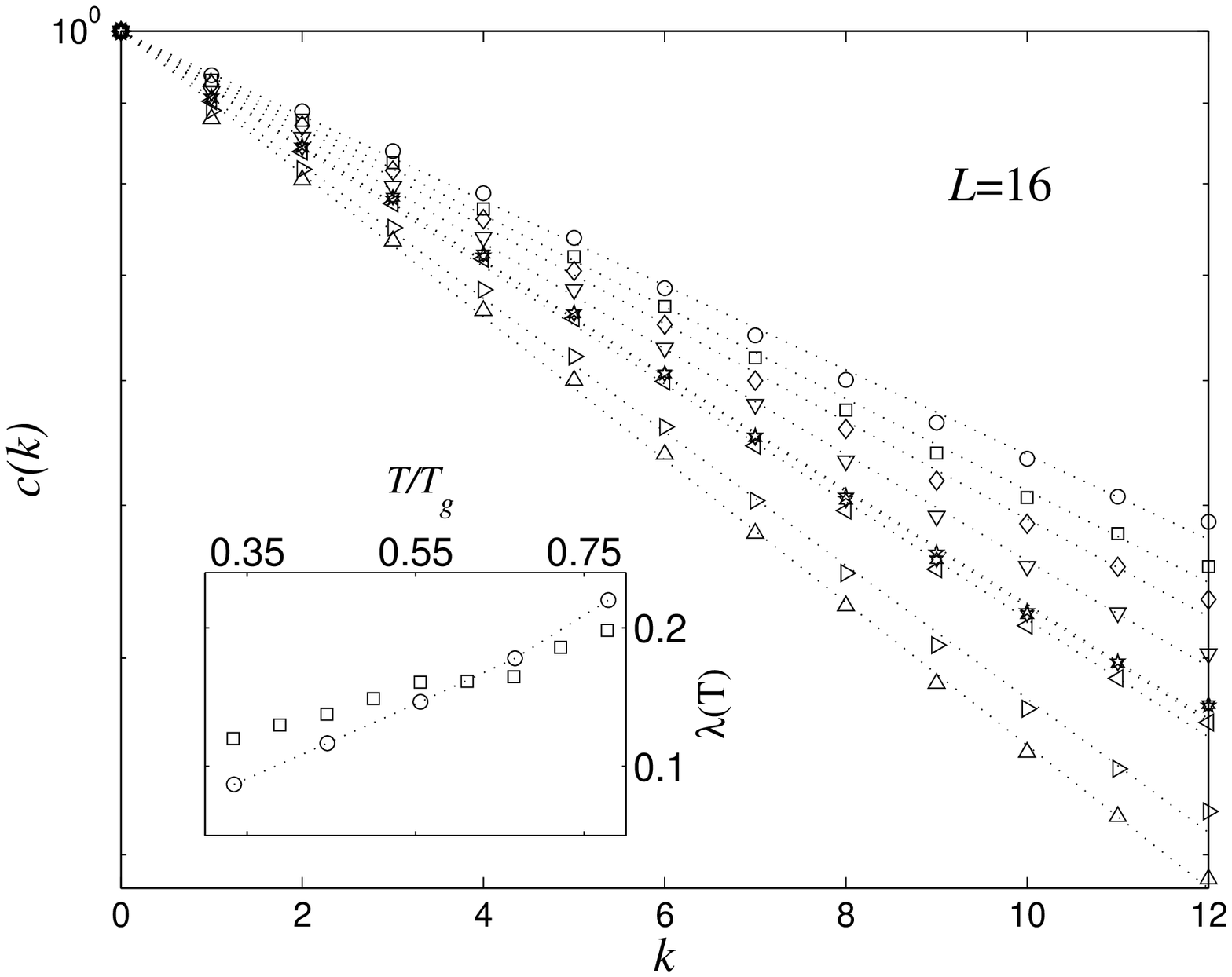}
\caption{ \small 
The average and variance of the number of valleys $ n_V$ visited
in time $[10,t_w]$. Their ratio is shown as a function of time in the insert. 
After a short transient, average and variance are nearly proportional 
to one another and to $\log t_w$, indicating that $n_V$ is proportional to 
a log-Poisson variable. 
In the time interval $[10^4,10^6]$, 
the logarithmic slope  is $\alpha = 1.63(1)$. 
}
\label{no_valleys} 
\caption{ \small 
The main panel shows that the assumed  exponential form of 
$(c(k)-c(\infty))/(1- c(\infty))$ versus $k$ is supported by the data,
for temperatures  $T=0.30$ (upper curve) $0.35, 0.40\dots 0.70$.
The lines are obtained by a  two-parameter fits, and the 
symbols correspond to data points.   
The insert shows the temperature dependence of the non-equilibrium
exponent  $\lambda(T,t_w=10)$ according to~\cite{Kisker96} (circles)
and as given by Eq.~(\ref{expos}) with $\alpha = 2.0$.   The age argument 
$t_w=10$ was chosen for the comparison, as it corresponds to 
 the earliest time at which our  data were collected. All temperatures
 are divided by $T_g = 0.9$. 
} 
\label{correlation}
\end{figure}

The set of data shown in Fig.~\ref{no_valleys} was obtained for  
a system of linear size $L=30$ at temperature $T=0.5$. 
Numerous similar plots (not shown) were obtained for 
temperatures in the range $0.2 \leq T  \leq 0.8$, for 
  $8\leq L \leq 30$ and $t_w \leq 10^6$.  
In most cases $20000$ independent runs were performed for each 
value of $T$ and $L$, removing any  visible  statistical flutter. When 
$\alpha$ is calculated as the logarithmic slope of the number of valleys
seen on average in the time window $[10^4,10^6]$, its size dependence 
for $L=13,16,20$ and $30$ is found to be
$\alpha=1.17(1)$, $1.37(1)$, $1.58(1)$ and $1.63(1)$, respectively. 
While the growth clearly tapers  off,  we cannot with confidence 
conclude on the true asymptotic behavior
for large $L$. We note, however, that   $\lambda(T)  \approx \alpha \gamma(T)$ should remain finite
in the thermodynamic limit. Hence, the growth of 
$\alpha$  must either  remain  bounded, 
 or be   compensated by a 
corresponding decrease of $\gamma(T)$. To see that the latter
is likely the  case, we note that  $\gamma = -\log(1 - 2 H_d/L^d)$, where $H_d$ 
 is  the Hamming distance   between the local minima at the bottom of  
consecutive ($k=1$) valleys. In 3 d, $H_d \propto L^{2.8}$ was found
 in~\cite{Dall03},  whence, using  
 $H_d/L^d <<1$  one obtains the slow decay  $\gamma \propto L^{-0.2}$. 

In order to obtain a   proper log-Poisson variable,  
 the  number of valleys must be divided by   $\approx 0.7$, the
ratio  between   variance  and average, and the 
  empirical value of $\alpha$ must be similarly rescaled  
 to the  effective value $\alpha \approx 2.0$ 
for a $16^3$ system, and $\alpha \approx 2.3$ for a $30^3$ system. 
 
Next, we analyze the distribution of the residence time $t_r$ within valleys 
entered at time $[0.95 t_w, 1.05 t_w]$ for $10^2 \leq t_w \leq 10^5$.  
The observed frequencies of the scaled quantities $t_r/t_w^\mu$ 
with $\mu = 1.05$ are obtained from the same runs as in Fig.~\ref{no_valleys}. 
They are plotted as lines in the main panel of Fig.~\ref{residence_t}, 
and the excellent quality of the resulting collapse gives the false impression 
of a single continuous line. The same distributions are also plotted in the insert, 
this time with $\mu=1$, which is the theoretical prediction of the log-Poisson 
statistics. We see that while $\mu = 1$ fails to produce a complete data collapse, 
it nevertheless accounts for the main features of the distribution. 
The corresponding theoretical prediction of Eq.~(\ref{an_scaling}) with 
$\alpha=2.3$ is given by the circles in the main panel.

We finally  consider the correlation decay.
The overlap $c(1;k+1)$---for short $c(k)$---between the lowest lying configurations
of the first and the $(1+k)$'th valley was calculated separately for temperatures 
$T=0.3,0.35,\ldots, 0.7$ starting in each case from a random initial condition. 
In this case, data are for each $T$ averages over $2000$ trajectories belonging  
to $20$ different realizations of the couplings. 
Since no data were taken for the first $10$ time steps, the smallest age value 
considered is $t_w=10$, rather than the theoretical value $t_w=1$. 
A two parameter non-linear fit to the exponential form in Eq.~(\ref{twoparameterfit}) 
was performed, producing the lines shown in the main panel of Fig.~\ref{correlation},
while the actual data are given by the symbols. The main error in fitting 
$c(k)$ is systematic and originates from the poor sampling of late events, 
which is dealt with by introducing a cut-off at $k = 12$. 
The fitted values of $\gamma$ and $c_\infty$ vary a few percent  as 
the position of the cut-off is varied slightly around this value.
The insert of Fig.~\ref{correlation} compares 
the temperature dependence of the predicted non-equilibrium exponent
$\lambda \approx \alpha \gamma(T)$ (squares) with the corresponding result by 
Kisker et al.~\cite{Kisker96}. The comparison is primarily qualitative, 
since we are unable to determine $\alpha$ with great precision. 
We used $\alpha = 2.0$ in this plot. 
  
The (implied) temperature dependence of $\gamma(T)$
shows that, for the same number of quakes, configurations decorrelate
faster at higher temperatures, as one would expect.
We note in passing that $c_\infty$ systematically decreases with $T$, 
and that $c_\infty$ would equal the Edwards-Anderson order parameter 
if the simple exponential decay of $c(k)$ were to continue indefinitely. 

While the exponential decorrelation form given by Eq.~(\ref{simple_decorr})---and 
the self-similar picture of the landscape it implies--- 
is required for full aging, it is is not likely 
to be exactly fulfilled on all time scales. Yet it  agrees with  
the full aging behavior observed in spin glasses~\cite{rodriguez02},
which highlights the   dynamical nature of their behavior. 
Interestingly, strong  superaging behavior is found in gels~\cite{Bissig03}.
Hence, either the latter systems 
are closer to thermal equilibrium than spin glasses,  or  
  considerable   deviations  from the marginal stability increase
of the visited attractors occur.

\acknowledgments 
This project has been supported by Statens Na\-tur\-viden\-skabe\-lige 
Forsk\-nings\-r\aa d through a block grant 
 and by the Danish Center for Super Computing with computer time
on the Horseshoe Linux Cluster. We are grateful to S. Boettcher
and to J.C. Sch\"{o}n for useful comments and discussions. 


\end{document}